\documentclass[12pt,twoside]{article}
\usepackage{fleqn,espcrc1}

\usepackage{graphicx}
\usepackage[figuresright]{rotating}

\newcommand{\AmS}{{\protect\the\textfont2
  A\kern-.1667em\lower.5ex\hbox{M}\kern-.125emS}}

\title{Possible $S-$wave Dibaryons in SU(3) Chiral Quark Model
            \thanks{This work is supported in part by the National Natural
                Science Foundation of China (NSFC) and the Chinese Academy
                Of Sciences.}}
\author{P.N.Shen$^{a,b}$, Q.B.Li$^a$, Z.Y.Zhang$^a$ and Y.W.Yu$^a$\\
\vspace{0.3cm}
$~^a$Institute of High Energy Physics, CAS, P.O.Box 918(4), Beijing 100039, 
         China\\
\vspace{0.3cm}
$~^b$CCAST (World Laboratory), P.O.Box 8370, Beijing 100080, China}
\begin{document}

% typeset front matter
\maketitle

\begin{abstract}
In the framework of the SU(3) chiral quark model, the $S-$wave
baryon-baryon bound states are investigated. It is found that according 
to the symmetry character of the system and 
the contributions from chiral fields, there are three types of bound states.
The states of the first type, such as $[\Omega\Omega]_{(0,0)}$
and $[\Xi^{*}\Omega]_{(0,1/2)}$ are deeply bound dibaryon with narrow widths.
The second type states, $[\Sigma^{*} \Delta]_{(0,5/2)}$,$[\Sigma^{*} \Delta]_{(3,1/2)}$, $[\Delta\Delta]_{(0,3)}$ and
$[\Delta\Delta]_{(3,0)}$ are also bound states, but with broad widths.
$[\Xi\Omega - \Xi^{*}\Omega]_{(1,1/2)}$, $[\Xi\Xi]_{(0,1)}$,
and $[N \Omega]_{(2,1/2)}$ are third type states. They, like {\em d}, are 
weakly bound only if the chiral fields can provide attraction
between baryons.

\end{abstract}

\section{Introduction}

In recent years, there is quite a bit activity on searching dibaryons
both theoretically and experimentally. Because it is commonly believed
that the radius of the dibaryon is less than $0.85fm$, studying dibaryon 
would enhance our knowledge about the short-range behavior of the
Quantum Chromodynamics (QCD), provide the information about the nonperturbative
QCD (NPQCD) and its appropriate treatment. Moreover, the existence of
dibaryon is a direct evidence of quarks in hadrons and hadronic systems.

Since Jaffe predicted H \cite{ja}, there have been lots of efforts 
devoted to the dibaryon study. The most intensive studied dibaryons are 
H, $D^*$ and $d^{\prime}$. Up to now, the most believed value of H is around
The $\Lambda\Lambda$ theshold \cite{ya,szy}, which agrees with the experimental
finding, the lower limit of $2220MeV$ \cite{im}. The theoretical 
prediction for $d^*$ varies model by model, but most believed value is
several tens of MeV \cite{wan,yzys}. The theoretical result of $d^{\prime}$ is
still away from the expected value extracted from double-charge-exchange
reactions \cite{wag}. Therefore, it is reasonable to propose studies on
the systems with multi-strangeness. With this idea, $\Omega\Omega(S=0,T=0)$
and $\Xi^{*}\Omega(S=0,T=1/2)$ were studied \cite{yzy,ls}. In this talk, we
would demonstrate a systematic study of the possible $S-$wave dibaryons.

\section{Model for studying dibaryon}

Because NPQCD still cannot directly be solved, one has to use model to 
approximately treat the NPQCD effect. In order to reliably study 
dibaryon, the employed model should at least satisfy two-fold requirements: 
(1) The ground state properties of baryons, and the available experimental 
data of the nucleon-nucleon and the nucleon-hyperon ($N-Y$) scattering 
can  reasonably be reproduced in the dynamical calculation. (2) In the 
extended application, no additional parameters are required.
The SU(3) chiral quark model is just one of the models which meets 
the requirements \cite{zys}.
% In this model, the short-range perturbative
%behavior of QCD is offered by the one-gluon-exchange interaction, the short- 
%and medium-range and the long-range NPQCD effects are provided by the 
%chiral-quark field interactions and the confinement potential, respectively.
The details of the model can be found in ref.\cite{zys}. 

The model parameters are fixed before the calculation \cite{zys}. The set of
values used in ref.\cite{zys}is called Model I (Set I).
In order to further reduce the number of free parameters, 
we choose the masses of strange scalar mesons, $\kappa$ and $\epsilon$,
to be the values of the mesons with the same quantum numbers in the 
particle data table. Moreover, in order to test the effects of various
chiral fields, we also establish the extended SU(2)chiral quark model 
(Model II), where $\sigma^{\prime}$, $\kappa$ and $\epsilon$ are disregarded,
and the SU(2) chiral quark model (Model III), where only $\pi$ and 
$\sigma$ are presented.

The bound state problem of the six-quark system is solved by using the 
Resonating Group Method (RGM). In RGM, the trial wave function of the system
can be written as 
\begin{equation}
\Psi={\cal A}[ \hat{\phi_A}(\xi_{1},\xi_{2}) \hat{\phi_B}(\xi_{4},\xi_{5})
 \chi_{rel} (\vec{R}) \chi_{CM}(\vec{R}_{CM})],
\end{equation}
where $\chi_{rel} (\vec{R})$ describes the trial wave function of 
of relative motion between baryon clusters $A$ and $B$ 
and  ${\cal A}$ is the antisymmetrizer. In terms of the variational method, 
one can solve $\chi_{rel} (\vec{R})$ and consequently the binding energy
of the system.

\section{Results and discussion}

As is well known, the matrix element $\langle {\cal A}^{sfc} \rangle$, where
the superscript $sfc$ implies that ${\cal A}$ acts in the $spin-flavor-color$
space only, is an important measure of the action of the Pauli principle 
on the two-baryon state. It also specifies the symmetry character of 
the system. The symmetry character of the two-baryon system could generally 
be divided into  three categoyies. In the first category, 
$ \langle {\cal A}^{sfc} \rangle\sim -2$.
This feature would be enormously beneficial to form a state with 
the $[6]_{orbit}$ symmetry. If the inter-baryon interaction further 
demonstrates the attractive character with certain strength, it is possible 
to form a smaller-sized bound state. 
In the second category, $\langle {\cal A}^{sfc} \rangle\sim 1$. Thus, the 
Pauli blocking effect here are very small so that two baryons are 
basically independent with each other. Then, if the inter-baryon 
meson-exchange interaction can provide large enough attraction, the system 
would be bound. This result may also be deduced in terms of a model in the 
baryon-meson degrees of freedom.
The states with $ \langle {\cal A}^{sfc} \rangle\sim 0$ belong to the third 
category. The Pauli blocking effect between baryons are incredibly serious 
so that the state with $[6]_{orbit}$ is almost a forbidden state, namely, it 
is very hard to form a bound state.

We present the calculated binding energies and the corresponding 
root-mean-square radii (${\cal R}$) for the possible $S-$wave 
states in the first and second categories in 
%Tables~\ref{\table:1}$(a)$ and $(b)$, 
Tables 1$(a)$ and $(b)$, respectively. 
\begin{table}[htb]
\label{table:1}
\newcommand{\m}{\hphantom{$-$}}
\newcommand{\cc}[1]{\multicolumn{1}{c}{#1}}
\renewcommand{\tabcolsep}{0.4pc} % enlarge column spacing
\renewcommand{\arraystretch}{1.2} % enlarge line spacing
\begin{footnotesize}
\begin{tabular}{@{}lllllll}
\multicolumn{7}{l}{\normalsize {\hspace{-0.4cm}Table 1}}\\
\multicolumn{7}{l}{\normalsize {\hspace{-0.4cm}(a) Binding energy $E_b$ and 
corresponding ${\cal R}$ for the systems with $\langle {\cal A}^{sfc}
\rangle \approx 2$.}}\\
\hline
{$s$} & System & $\langle {\cal A}^{sfc} \rangle$ & Model I (Set I) & 
Model I (set II) & Model II & Model III\\
  &  &  & $E_b~~//{\cal R}$ & $E_b~~//{\cal R}$ 
  & $E_b~~//{\cal R}$ & $E_b~~//{\cal R}$ \\ 
\hline
  0 & $\Delta\Delta(S=3,T=0)~(d^{*})$ & $2$ & 22.2~//1.01 
     & 18.5~//1.05 & 64.8//0.84 & 62.7//0.86 \\  
  & $\Delta\Delta (S=0,T=3)$ & $2$ & 16.0~//1.10 
     & 13.5~//1.14 &  6.3//1.25 & 13.2//1.11 \\ 
 -1 & $\Sigma^* \Delta (S=0,T=5/2)$ & $2$
     & 24.6~//0.99 & 19.0~//1.04 & 11.7//1.13 & 21.5//0.99 \\  
    & $\Sigma^* \Delta (S=3,T=1/2)$ & $2$
     & 25.9~//0.95 & 29.3~//0.93 & 78.0//0.79 & 76.3//0.80 \\  
 -5 & $\Xi^* \Omega(S=0,T=1/2)$ & $2$
     & 92.4~//0.71 & 76.5~//0.72 & 58.8//0.75 & 52.1//0.76 \\ 
 -6 & $\Omega\Omega(S=0,T=0)$ & $2$
     & 113.8~//0.66 & 98.5~//0.67 &  72.6//0.70 &  51.9//0.73 \\ 
\hline \hline    
\multicolumn{7}{l}{}\\
\multicolumn{7}{l}{\normalsize {\hspace{-0.4cm}(b) Binding energy $E_b$ and 
corresponding ${\cal R}$ for the systems with $\langle {\cal A}^{sfc}
\rangle \approx 1$.}}\\
\hline
{$s$} & System & $\langle {\cal A}^{sfc} \rangle$ & Model I (Set I) & 
Model I (set II) & Model II & Model III\\
  &  &  & $E_b~~//{\cal R}$ & $E_b~~//{\cal R}$ 
  & $E_b~~//{\cal R}$ & $E_b~~//{\cal R}$ \\ 
\hline
%-2 & $H$ particle &    & -2.0~//1.15 & 8.2~//0.91 \\ 
-3 & $N\Omega(S=2,T=1/2)      $ & 1 & 3.5~//1.18 & 12.7~//0.98
    &  31.8~//0.81  &  49.5~//0.74 \\ 
& $\Delta\Omega (S=3,T=3/2)$ & 1 & 4.4~//1.15 & 14.2~//0.96
    & 34.3~//0.80 & 49.6~//0.74 \\  
-4  & $\Xi\Xi(S=0,T=1)$ & $10/9$ & 4.1~//1.17  & 0.4~//1.30
    & -0.5~//1.33 & 3.1~//1.18 \\ 
-5 & $\Xi\Omega-\Xi^*\Omega(S=1,T=1/2)$ & & 32.9~//0.78 & 32.6~//0.77
    & 29.3~//0.79 & 17.6~//0.86 \\ 
\hline    
\end{tabular}\\[2pt] 
The units for $E_b$ and RMS are in $MeV$ and $fm$, respectively. {\em s}
Stands for the strangeness.
\end{footnotesize}
\end{table}
\noindent

In order to see the effects of various chiral fields and give a possible 
range of predicted bound states, we also study these states in terms of
Models II and III and tabulated the result in Table 1.
From Table 1$(a)$, one finds that in the $s=-6$ and $-5$ cases,
The systems in Model I are bounder than in Models II and III, because the
strong attraction provided by $\sigma$ and strange mesons $\kappa$ and 
$\epsilon$. If one believes that the SU(3) model is more suitable for the
multi-strangeness system, $[\Omega\Omega]_{(0,0)}$ and
$[\Xi^{*}\Omega]_{(0,1/2)}$ should be deeply bound states with narrow widths.
On the other hand, in the $s=-1$ and $0$ systems, Models I and II
are favored by the low-spin and high-spin systems, respectively. 
Because the color-magnetic interaction (CMI) shows repulsive and attractive
features in the $S=0$ and $3$ systems, respectively, the 
binding energies of the $S=3$ systems are generally higher than those of 
the $S=0$ systems. Moreover, due to the strong decay modes of both composed
baryons, the widths of these bound states are broad.
The results in Table 1$(b)$ shows another type of character. 
Because, like $d$, the exchange effect is almost negligible, the resultant 
binding energy is very sensitive to the contributions
from chiral fields. It is shown that the high-strangeness system is bounder
in Model I and the low-strangeness system has larger binding energy
in Model III. Nevertheless, these systems are weakly bound.

\begin{figure}[htb]
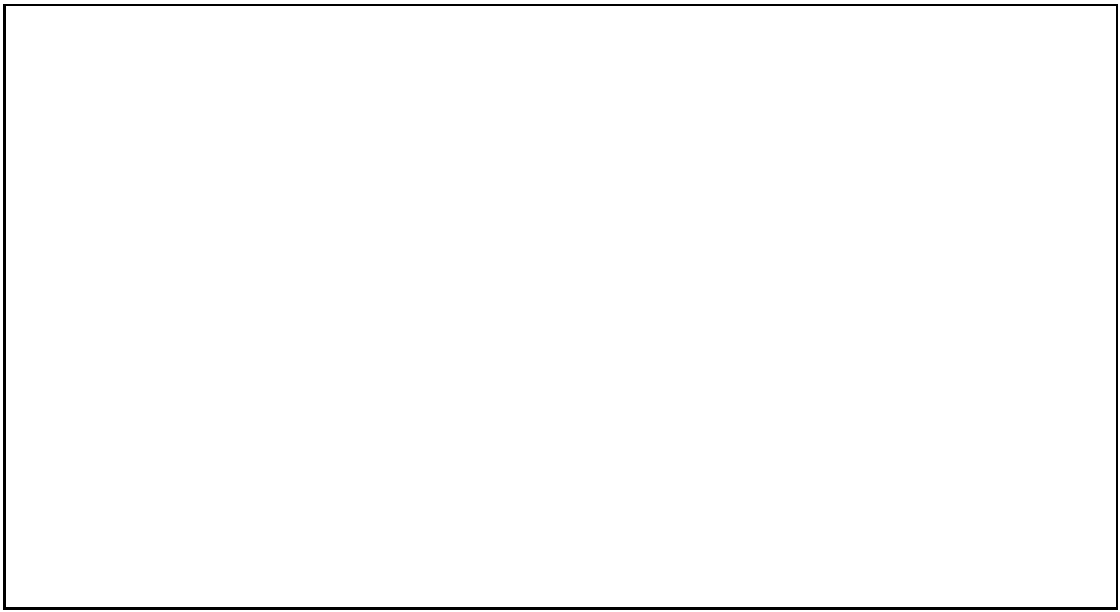

\begin{minipage}[t]{150mm}
\framebox[148mm]{\rule[-26mm]{0mm}{78mm}}
\caption{Binding energy change with respect to $b_u$ and $m_s$.}
\label{fig:1}
\end{minipage}
\end{figure}

Finally, to ensure the predicted bound states are reliable and stable, we also
examine their model-parameter dependencies. The $m_{\kappa}$- and
$m_{\epsilon}$-dependencies are demonstrated in Table 1 by Sets I and II. 
It is shown that the qualitative binding behavior does not change although
the binding of most systems become slightly weaker due to the overall weaker
attractive feature from $\kappa$ and $\epsilon$. Moreover, the $m_s$-
and $b_u$-dependencies are plotted in Fig.~\ref{fig:1}. It is shown that 
the binding energy always becomes smaller with increasing $b_u$ and decreasing 
$m_s$, but the qualitative binding behavior does not change. Therefore, our 
predicted bound states are stable and reliable.

\section{Conclusion}
We investigate the possible $S-$wave baryon-baryon bound states with the 
SU(3) chiral quark model, and test the reliabilities and stabilities of 
Results by using different models and model parameters. It is shown that the 
symmetry property, characterized by the expectation value of antisymmetrizer,
of the system is a major factor to affect the binding behavior. 
Sometimes, the contributions from various chiral fields demonstrate
very important roles. We arrange all the bound states into three types. 
The first type states, such as $[\Omega\Omega]_{(0,0)}$ and
$[\Xi^{*}\Omega]_{(0,1/2)}$ should be deeply bound states with narrow 
widths, or dibaryons.The states of second type, such as
$[\Sigma^{*} \Delta]_{(0,5/2)}$, $[\Sigma^{*} \Delta]_{(3,1/2)}$,
$[\Delta\Delta]_{(0,3)}$ and $[\Delta\Delta]_{(3,0)}~(d^{*})$ are bound states
with broad widths. The states like $[\Xi\Omega - \Xi^{*}\Omega]_{(1,1/2)}$,
$[\Xi\Xi]_{(0,1)}$, and $[N \Omega]_{(2,1/2)}$ are third types states. 
Like $d$, they generally the weakly bound states, and their binding 
behaviors very much depend on the contributions from the chiral fields. 
We would call great attention from physicists on the states of the first 
and second types and wish that these states can be searched in the future 
experiment.

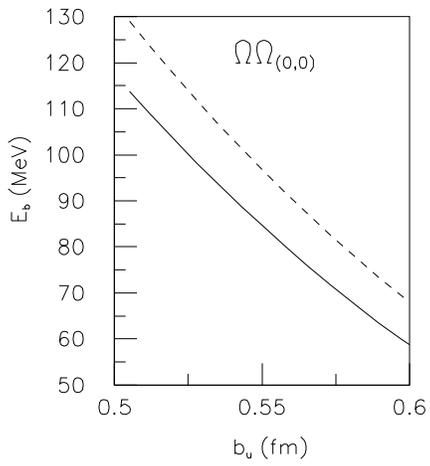
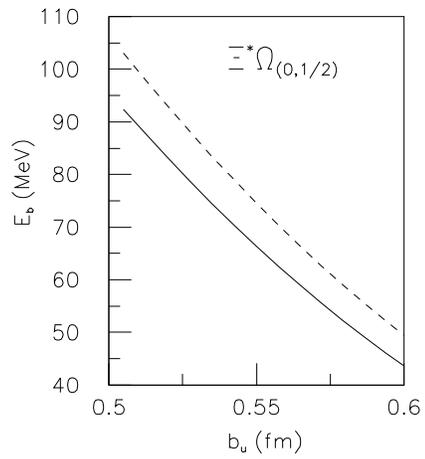
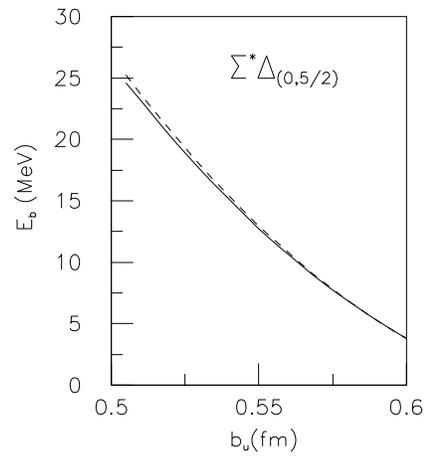
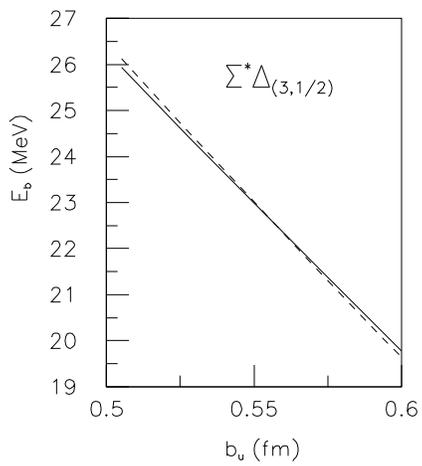
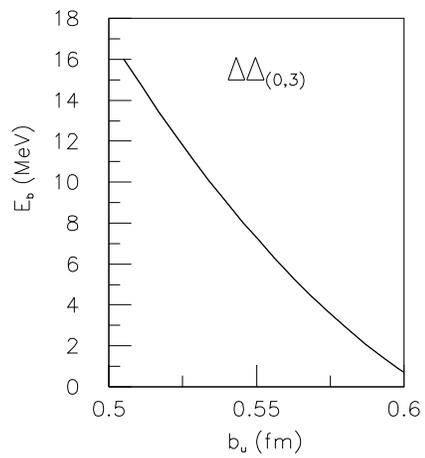
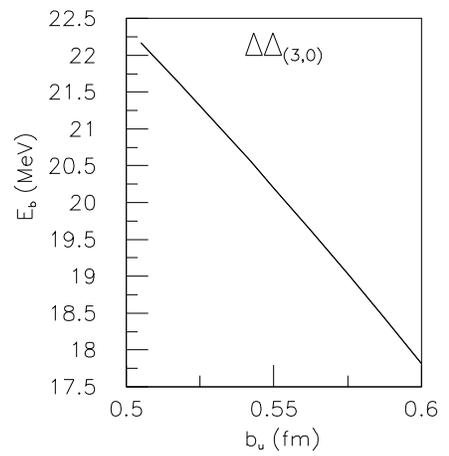
\end{document}